\documentclass[letterpaper, 10 pt, conference]{ieeeconf}  

\IEEEoverridecommandlockouts                              
\usepackage{epsfig} 
\usepackage{times} 
\usepackage{amsmath} 
\usepackage{amssymb}  
\usepackage{mathtools}
\usepackage{ntheorem}
\usepackage{arydshln}
\usepackage{comment}
\usepackage{xcolor}

\newtheorem{definition}{Definition}

\newtheorem{lemma}{Lemma}

\DeclareMathOperator*{\argmin}{arg\,min}

\DeclareMathOperator*{\diag}{diag}
\DeclareMathOperator*{\trace}{tr}

\usepackage{soul}

\newcommand{\vim}{v_{i-1}}
\newcommand{\aim}{a_{i-1}}

\newcommand{\dvim}{\dot{v}_{i-1}}
\newcommand{\daim}{\dot{a}_{i-1}}
\newcommand{\uim}{u_{i-1}}
\newcommand{\imo}{_{i-1}}

\makeatletter
\renewcommand{\section}{\@startsection{section}{1}{\z@}{1.5ex plus 1.5ex minus 0.5ex}%
	{0.7ex plus 1ex minus 0ex}{\normalfont\normalsize\centering\scshape}}%
\renewcommand{\subsection}{\@startsection{subsection}{2}{\z@}{1.5ex plus 1.5ex minus 0.5ex}%
	{0.7ex plus 1ex minus 0ex}{\normalfont\normalsize\itshape}}%
\makeatother

\expandafter\def\expandafter\normalsize\expandafter{%
\normalsize
\setlength\abovedisplayskip{5pt}
\setlength\belowdisplayskip{5pt}
\setlength\abovedisplayshortskip{5pt}
\setlength\belowdisplayshortskip{5pt}
}

\newcommand{\runScript}[2]{#2}

\title{\LARGE \bf
Optimal Controller Realizations against False Data Injections in Cooperative Driving}

\author{Mischa Huisman, Carlos Murguia, Erjen Lefeber, and Nathan van de Wouw
\thanks{The research leading to these results has received funding from the European Union’s Horizon Europe programme under grant agreement No 101069748 – SELFY project.}
\thanks{M. Huisman, C. Murguia, E. Lefeber, and N. van de Wouw are with the Department of Mechanical Engineering, Eindhoven University of Technology, The Netherlands.
        {\tt\small [m.r.huisman}, {\tt\small C.G.Murguia}, {\tt\small A.A.J.Lefeber}, {\tt\small N.v.d.Wouw]@tue.nl}}
}

\begin{document}

\maketitle
\thispagestyle{empty}
\pagestyle{empty}

\begin{abstract}
To enhance the robustness of cooperative driving to cyberattacks, we study a controller-oriented approach to mitigate the effect of a class of False-Data Injection (FDI) attacks. By reformulating a given dynamic Cooperative Adaptive Cruise Control scheme (the base controller), we show that a class of new but equivalent controllers (base controller realizations) can represent the base controller. This controller class exhibits the same platooning behavior in the absence of attacks, but in the presence of attacks, their robustness varies with the realization. We propose a prescriptive synthesis framework where the base controller and the system dynamics are written in new coordinates via an invertible coordinate transformation on the controller state. Because the input-output behavior is invariant under coordinate transformations, the input-output behavior is unaffected (so controller realizations do not change the system's closed-loop performance). However, each controller realization may require a different combination of sensors. Subsequently, we obtain the optimal combination of sensors that minimizes the effect of FDI attacks by solving a linear matrix inequality while quantifying the FDI's attack impact through reachability analysis. Through simulation studies, we demonstrate that this approach enhances the robustness of cooperative driving without relying on a detection scheme and maintaining all system properties.


\end{abstract}

\section{INTRODUCTION} \vspace{-1mm}
Cooperative Adaptive Cruise Control (CACC) is a well-explored technology within Connected and Automated Vehicles (CAVs) that allows groups of vehicles to form tightly-coupled platoons by exchanging inter-vehicle data through Vehicle-to-Vehicle (V2V) wireless communication networks \cite{ploeg_design_2011}-\cite{lefeber_cooperative_2020}. However, as CACC requires communication networks, network access points are exposed that adversaries can exploit for cyberattacks \cite{amoozadeh_security_2015}-\cite{ sun_survey_2022}. To counter these attacks, technologies are being developed to prevent and detect cyberattacks, increasing system security \cite{ju_survey_2022}-\cite{chowdhury_attacks_2020}. However, the effectiveness of current mitigation technology is limited by: (i) unknown process or measurement disturbances \cite{teixeira_attack_2012, zhang_networked_2019}, (ii) limited operating resources (e.g., computing power, budget) \cite{anand_risk_2022}, and (iii) adversaries exploiting model knowledge \cite{anand_risk_2022,teixeira_strategic_2015}. Therefore, technology is needed to enhance the robustness of cooperative driving to attacks on system elements such as sensors, actuators, networks, and software. 

To address these challenges, the literature offers two control-theoretic approaches: (i) active methods that leverage attack detection to switch controllers and (ii) passive attack-resilient methods that seek to withstand the effect of attacks. \\ The\hspace{-0.3mm} former\hspace{-0.3mm} relies\hspace{-0.3mm} on\hspace{-0.3mm} a\hspace{-0.3mm} detection\hspace{-0.3mm} scheme\hspace{-0.3mm} to\hspace{-0.3mm} notify\hspace{-0.3mm} the\hspace{-0.3mm} controller\hspace{-0.3mm} to\hspace{-0.3mm} switch\hspace{-0.3mm} towards\hspace{-0.3mm} a\hspace{-0.3mm} fallback\hspace{-0.3mm} mechanism\hspace{-0.3mm} (e.g.,\hspace{-0.3mm} from\hspace{-0.3mm} CACC\hspace{-0.3mm} to\hspace{-0.3mm} adaptive\hspace{-0.3mm} cruise\hspace{-0.3mm} control\hspace{-0.3mm} \cite{van_der_heijden_analyzing_2017}), while the latter often compromises the nominal performance (e.g., string stability \cite{wolf_securing_2020}).  

In this paper, we introduce a third controller-oriented approach to implement a given dynamic CACC scheme without affecting the input-output behavior of the closed-loop system, thereby preserving nominal performance. By reformulating the dynamic CACC scheme (the base controller), we show that the base controller can be represented by a class of equivalent realizations (base controller realizations). These realizations have equivalent nominal behavior with varying robustness in the presence of attacks. This approach enhances the robustness of cooperative driving without relying on a detection scheme and maintains system properties such as string stability. 

We demonstrate that a different controller realization may require a different combination of sensors from a set of sensors, which could be subject to a resource-limited False Data Injection (FDI) attack. The effect of such an FDI attack can be quantified using reachability analysis to obtain the adversarial reachable set, which provides insight into the size of the state space portion that the FDI attack can induce \cite{murguia_security_2020}. Moreover, the reachable set provides insight into the potential damage an FDI attack can do to a platoon, e.g., cause a collision. To this end, we formulate a Semi-Definite Program (SDP) to obtain the optimal controller realization that minimizes the size of the adversarial reachable set. Moreover, the corresponding synthesis problem results in a Linear Matrix Inequality (LMI). A simulation study is conducted to support our claims, where the original CACC realization, an alternative formulation, and the optimal realization are compared. Our findings reveal that the optimal realization has the smallest size of the reachable set within the considered class of controllers. 

The structure of the paper is as follows. Section~\ref{sec:Preliminaries} introduces preliminary results. Section~\ref{sec:ProblemSetting} describes the general problem setting, including the platooning dynamics, the dynamic CACC scheme, and the available measurements. Section~\ref{sec:Approach} presents the class of controllers that can be derived from the given problem setting, after which the optimization problem is introduced. Section~\ref{sec:Results} presents the optimal realization, and a simulation study is conducted to demonstrate its performance. Finally, Section~\ref{sec:Conclusion} provides the concluding remarks.

\vspace{-1mm} \section{Mathematical Preliminaries} \vspace{-1mm} \label{sec:Preliminaries}

\subsection{Notation} \vspace{-1mm}
The symbol $\mathbb{R}$ stands for the real numbers, $\mathbb{R}_{>0}$ ($\mathbb{R}_{\geq 0}$) denotes the set of positive (non-negative) real numbers. The symbol $\mathbb{N}$ denotes the set of natural numbers, including zero. The $n \! \times \! m$ matrix composed of only zeros is denoted by $\mathbf{0}_{n \times m}$, or $\mathbf{0}$ when its dimension is clear. Consider a finite index set $\mathcal{L} \! \coloneqq \! \{l_1,\ldots,l_\rho\}  \! \! \subset \! \! \mathbb{N}$, then $\! \diag[B_j]$ and $(B_j),  j \! \in \! \mathcal{L}$, stand for the diagonal block matrix $\diag[B_{l_1},\ldots,B_{l_{\rho}}]$ and stacked block matrix $(B_{l_1},\ldots,B_{l_{\rho}})$, respectively. The notation $A \geq 0$ (resp., $A \leq 0$) indicates that the matrix $A$ is positive (resp., negative) semidefinite, i.e., all the eigenvalues of the symmetric matrix $A$ are positive (resp. negative) or equal to zero, whereas $A \! \! > \! \! 0$ (resp., $(A \! \! < \! \! 0)$) indicates the positive (resp., negative) definiteness, i.e., all the eigenvalues are strictly positive (resp. negative). Time dependencies of signals are often omitted for simplicity of notation.

\subsection{Definitions and Preliminary Results} \vspace{-1mm}
\begin{definition}[Reachable Set]\emph{\cite{murguia_security_2020}}
Consider the perturbed Linear Time-Invariant (LTI) system:\label{def1}
\begin{equation}
    \label{eq:LTI_set}
    \zeta(k+1) = \mathcal{A} \zeta(k) + \sum_{i=1}^N \mathcal{B}_i \omega_i(k), \ \zeta(0) = \zeta_0,
\end{equation}
with $k\in\mathbb{N}$, state $\zeta(k) \in \mathbb{R}^{n_\zeta}$, perturbation $\omega_i \in \mathbb{R}^{p_i}$ satisfying $\omega_i(k)^{\top} W_i \omega_i(k) \leq 1$ for some positive definite matrix $W_i \in \mathbb{R}^{p_i \times p_i}, i = \{1,...,N\}, N\in \mathbb{N}$, and matrices $\mathcal{A} \in \mathbb{R}^{n_\zeta \times n_\zeta}$ and $\mathcal{B}_i \in \mathbb{R}^{n_\zeta \times p_i}$. The reachable set $\mathcal{R}^{\zeta_0}$(k) at time $k\geq0$ from the initial condition $\zeta_0 \in \mathbb{R}^{n_\zeta}$ is the set of states reachable in $k$ steps by system \eqref{eq:LTI_set} through all possible disturbances satisfying $\omega_i^{\top}(k) W_i \omega_i(k) \leq 1$, i.e.,
\begin{equation}
    \! \! \mathcal{R}^{\zeta_0}(k) \! \coloneqq \!\!\left\{ \zeta \in \mathbb{R}^{n_\zeta} \left| \begin{array}{l}
        \zeta=\zeta(k), \zeta(k) \text{ \textit{solution to \eqref{eq:LTI_set},}}\\
         \text { \textit{and} } \omega_i(k)^{\top} W_i \omega_i(k) \leq 1.
    \end{array} \!\!\!\!  \right\} \right. \! .
\end{equation}
\end{definition}

\vspace{-1mm}
\begin{lemma}[Ellipsoidal Approximation]\emph{\cite{murguia_security_2020}}\label{lemma1}
	Consider the perturbed LTI system \eqref{eq:LTI_set} and the reachable set $\mathcal{R}^{\zeta_0}(k)$ in Definition \ref{def1}. For a given $a\in(0,1)$, if there exist constants $a_{1}$, $\ldots$, $a_{N}$ and matrix $P$ that is the solution of the convex program: \vspace{-1mm}
	\begin{equation}\label{lmi1}
		\left\{\!\begin{aligned}
            &\min_{P,a_{1},\ldots, a_{N}}-\log\det[P],\\
			&\text{s.t.} \hspace{1mm}a_{1},\ldots, a_{N}\in(0,1), \hspace{.5mm} a_{1}+\ldots+ a_{N}\geq a,\\
			&P>0, \begin{bmatrix}
				aP&\mathcal{A}^{\top}P&\mathbf{0}\\
				P\mathcal{A}&P&P\mathcal{B}\\
				\mathbf{0}&\mathcal{B}^{\top}P&W_{a}
			\end{bmatrix}\geq 0
		\end{aligned}\right.
	\end{equation}
with matrices $W_{a}:=\diag[(1-a_{1})W_{1},\ldots,(1-a_{N})W_{N}] \in \mathbb{R}^{\bar{p}\times\bar{p}}$, $\mathcal{B}:=(\mathcal{B}_{1},\ldots,\mathcal{B}_{N})\in\mathbb{R}^{n_{\zeta_0}\times \bar{p}}$, and $\bar{p}=\sum_{i=1}^{N}p_{i}$, then for all $k \in \mathbb{N}$, $\mathcal{R}^{\zeta_0}(k)\subseteq \mathcal{E}^{\zeta_0}(k)$ with $\mathcal{E}^{\zeta_0}(k):=\{ \zeta^{\top}(k)P^{\zeta_0}\zeta(k) \leq\alpha^{\zeta_0}(k)\} $, with convergent scalar sequence $\alpha^{\zeta_0}(k):=a^{k-1}\zeta(k)^{\top}P^{\zeta_0}\zeta(k)+((N-a)(1-a^{k-1}))/(1-a)$. Ellipsoid $\mathcal{E}^{\zeta_0}(k)$ has the minimum asymptotic volume among all outer ellipsoidal approximations of $\mathcal{R}^{\zeta_0}(k)$.
\end{lemma}

\vspace{-2mm}
\begin{lemma}[Projection]\emph{\cite{murguia_security_2020}}\label{lemma2}
	Consider the ellipsoid:
	\begin{equation}
		\begin{split}
			\mathcal{E}:=\left\lbrace x\in\mathbb{R}^{n}, y\in\mathbb{R}^{m}\bigg|\begin{bmatrix}
				x\\y
			\end{bmatrix}^{\top}\begin{bmatrix}
				Q_{1}& \hspace{-1mm}Q_{2}\\Q_{2}^{\top}&\hspace{-1mm}Q_{3}
			\end{bmatrix}\begin{bmatrix}
				x\\y
			\end{bmatrix}=\alpha \right\rbrace ,
		\end{split}
	\end{equation}
	for some positive definite matrix $Q\in\mathbb{R}^{(n+m)\times(n+m)}$ and constant $\alpha\in\mathbb{R}_{>0}$. The projection $\mathcal{E}'$ of $\mathcal{E}$ onto the $x$-hyperplane is given by the ellipsoid:
	\begin{equation}
		\mathcal{E}':=\left\lbrace x\in \mathbb{R}^{n} \left| \hspace{1mm} x^{\top}\begin{bmatrix}
			Q_{1}-Q_{2}Q_{3}^{-1}Q_{2}^{\top} 
		\end{bmatrix}x=\alpha\right\rbrace . \right.
	\end{equation}
\end{lemma}


\section{Problem Setting} \label{sec:ProblemSetting}
\begin{figure}[bt]\centering
		\includegraphics[width=\linewidth]{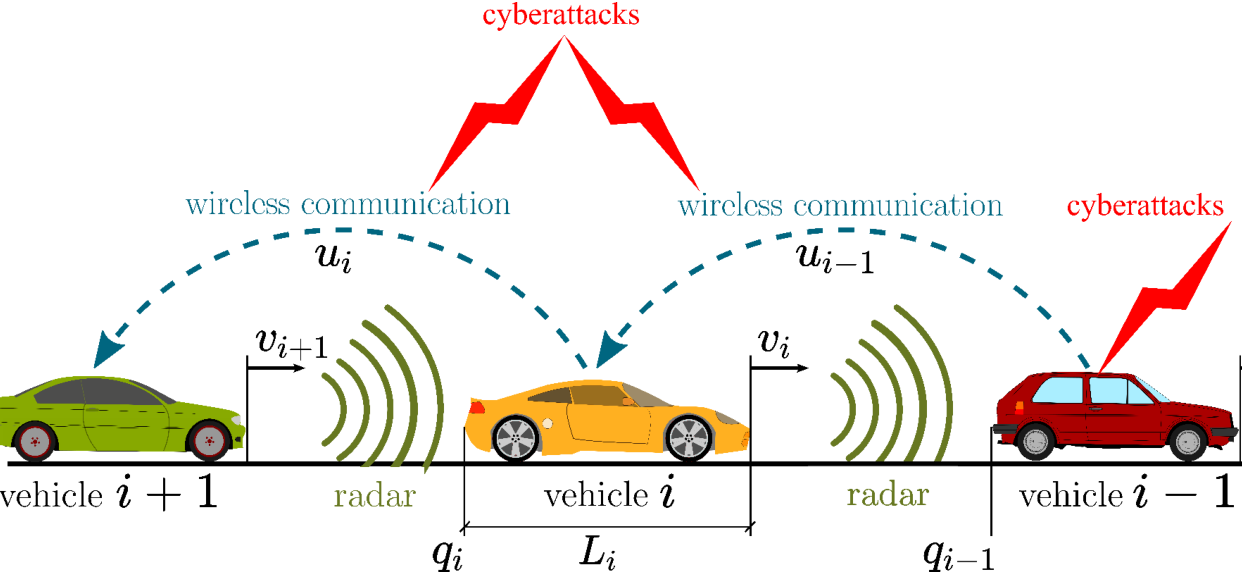}
		\caption{CACC-equipped vehicle platoon. Each vehicle has onboard sensors (e.g., radars/LiDARs, cameras, and velocity/acceleration sensors). Vehicles may be subject to FDI attacks.}
		\centering
		\label{fig:Platoon}
  \vspace{-6mm}
\end{figure}

Consider a homogeneous platoon of $m$ vehicles, schematically depicted in Fig. \ref{fig:Platoon}, where the vehicles are enumerated with index $i=1,...,m,$ with $i=1$ indicating the lead vehicle. To model such a platoon, we adopt the longitudinal vehicle model from \cite{ploeg_graceful_2013}:
\begin{align} \label{eq:SD_PlatoonDynamics}
    \begin{bmatrix}
        \dot{d}_i \\ \dot{v}_i \\\dot{a}_i
    \end{bmatrix}
    = 
    \begin{bmatrix}
        \vim - v_i \\ a_i \\ -\frac{1}{\tau}a_i + \frac{1}{\tau} u_i
    \end{bmatrix}, \, i \in S_m \backslash \{1\},
\end{align}
where $d_i = q_{i-1} - q_i - L_i$ ($q_i$ reflects the position of the rear bumper of vehicle $i$ and $L_i$ its length) being the distance between vehicle $i$ and its preceding vehicle $i-1$, $v_i$, and $a_i$ denoting the velocity, and acceleration of vehicle $i$, respectively, and $S_m \coloneqq \{ i \in \mathbb{N} \mid 1 \leq i \leq m \}$ (i.e., the set of all vehicles in a platoon of length $m \in \mathbb{N}$). The desired acceleration $u_i$ represents the control input, and $\tau>0$ is a time constant modeling driveline dynamics.

The objective of each follower vehicle is to keep a desired distance $d_{r,i}$ (the so-called constant time-gap policy) with its preceding vehicle:
\begin{align}
    d_{r,i} = r + h v_i, \, i \in S_m \backslash \{ 1 \}
\end{align}
with the time gap $h>0$, and standstill distance $r>0$. The spacing error is then defined as 
\begin{align}
\label{eq:SD_ErrorFormulation}
    \begin{split}
        e_i  &\coloneqq d_i - d_{r,i}.  
    \end{split}
\end{align}
To obtain the desired error dynamics, in \cite{ploeg_design_2011} a CACC controller is introduced to achieve string-stable vehicle-following behavior at small inter-vehicle distances for a homogeneous platoon. This CACC controller is a dynamic controller of the following form:
\begin{align} 
\label{eq:SD_BaseController}
        \mathcal{C} &\coloneqq \left\{ 
            \begin{aligned}
                u_i &= \rho_i, \\
                \dot{\rho}_i &= -\tfrac{1}{h} \rho_i + \tfrac{1}{h}(k_p e_i + k_d \dot{e}_i) + \tfrac{1}{h} \uim,
            \end{aligned} \right.           
\end{align}
which stabilizes the resulting error dynamics in \eqref{eq:SD_ErrorFormulation}. The constants $k_p>0$ and $k_d>0$ are control gains to be designed. An alternative to \eqref{eq:SD_BaseController} is the CACC controller introduced in \cite{lefeber_cooperative_2020}, which results in the same control signal $u_i$ at the vehicle but is now also applicable for heterogeneous platoons. This CACC controller is also dynamic and of the following form:
\begin{align} \label{eq:C2_Lefeber}
    \hat{\mathcal{C}} &\coloneqq \left\{ 
        \begin{aligned}
            \hat{u}_i &= -\tfrac{\tau}{h} \hat{\rho}_i + (1 - \tfrac{\tau}{h}) a_i + \tfrac{\tau}{h} \aim, \\
            \dot{\hat{\rho}}_i &= -\tfrac{1}{\tau} \hat{\rho}_i - \tfrac{1}{\tau_i}(k_p e_i + k_d \dot{e}_i).
        \end{aligned} \right.
\end{align}

Note that the real-time realization of any control scheme depends on the available sensors $y_{i,j}$ (sensor number $j$ of vehicle $i$). Considering a combination of sensor data coming from onboard sensors (e.g., LiDAR, radar, cameras, and velocity/acceleration sensors) and wirelessly received data from adjacent vehicles, we assume to have the following sensors to realize controllers $\mathcal{C}$ and $\hat{\mathcal{C}}$:
\begin{align} \label{eq:SD_Sensors}
    \begin{aligned}
        y_{i,1} &\coloneqq d_i + \delta_{i,1}, 
        \ \ \ \ \ \ \ \ y_{i,2} \coloneqq v_i + \delta_{i,2}, \\
        y_{i,3} &\coloneqq a_i + \delta_{i,3}, 
        \ \ \ \ \ \ \ \ y_{i,4} \coloneqq v_{i-1} - v_i + \delta_{i,4}, \\
        y_{i,5} &\coloneqq a_{i-1} + \delta_{i,5}, 
        \ \ \ \ \ y_{i,6} \coloneqq u_{i-1} + \delta_{i,6}.
    \end{aligned}
\end{align}
Herein, $\delta_{i,j}$ models potential FDI attacks. Sensors $y_{i,1}$ and $y_{i,4}$ provide relative distance and velocity information, $y_{i,2}$ and $y_{i,3}$ denote the onboard measured velocity and acceleration, while $y_{i,5}$ and $y_{i,6}$ model data received from the preceding vehicle via V2V communication. Using \eqref{eq:SD_Sensors}, it can be observed that a difference in realization between $\mathcal{C}$ and $\hat{\mathcal{C}}$ is the use of sensors $y_{i,6}$ and $y_{i,5}$, respectively. The findings in \cite{huisman_impact_2023} show that both controller realizations yield the same control input $u_i$ when $\delta_{i,j} = 0$; however, this equivalence is not guaranteed when $\delta_{i,j} \neq 0$. 

By applying a linear coordinate transformation to the internal state of the dynamic controller \eqref{eq:SD_BaseController}, also known as a similarity transformation, infinitely many real-time realizations of \eqref{eq:SD_BaseController} exist using sensors from \eqref{eq:SD_Sensors} ($\hat{\mathcal{C}}$ is just one of these realizations). Therefore, the problem we address in this paper is determining the optimal controller realization for \eqref{eq:SD_BaseController}, which minimizes the impact of resource-limited adversaries. The effect of a resource-limited FDI attack can be quantified using the size of the reachable set induced by such an attack. By minimizing the size of the adversarial reachable set, we select an optimal combination of sensors and the corresponding controller realizations to mitigate the effect of a potential FDI attack on the closed-loop dynamics.

\section{Approach for Optimal Controller Realization} \label{sec:Approach}
\runScript{
\subsection{Controller Realization} 
One could interpret the difference between the CACC controllers in \eqref{eq:SD_BaseController} and \eqref{eq:C2_Lefeber} as a coordinate transformation on the internal control variable $\rho_i$. Therefore, we propose to find the class of controllers that yields the same control signal $u_i$ while using the system output $y_i$ by applying the following coordinate transformation 
\begin{align} \label{eq:CR_Transformation}
  \bar{\rho}_i &= \alpha_i \rho_i + \underbrace{\begin{bmatrix} \beta_{i,1} & \beta_{i,2} & \beta_{i,3} & \beta_{i,4} & \beta_{i,5} & 0 \end{bmatrix}}_{\beta_i} y_i
\end{align}
with new controller state $\bar{\rho}_i$. The goal is to determine the optimal values for both $\alpha_i \in \mathbb{R}$ and $\beta_i\in \mathbb{R}^{n_y}$ ($n_y$ number of measurements) to reduce the impact of FDI attacks. Note that in the proposed change of coordinates $\beta_{i,6} = 0$ (excluding $y_{i,6}$), as this would require information about $\dot{u}\imo$, and thus also knowledge about the control structure of vehicle $i-1$ (and possibly sensor data from vehicle $i-2$).

The closed-loop system formulation is adopted from \cite{lefeber_cooperative_2020}, where it was observed that the resulting closed-loop dynamics, with output $e_i$ and input $u_i$, has the relative degree two. Therefore, we differentiate the output $e_i$ twice and then investigate the resulting internal dynamics. To this end, we define the stacked state vector $x = [ e_i \ \dot{e}_i \ z_i \ \vim \ \aim ]^\top$, where $z_i \! \coloneqq \! \vim - v_i$ is the internal dynamic state. Using \eqref{eq:SD_PlatoonDynamics} and \eqref{eq:SD_ErrorFormulation} we obtain the platooning dynamics of the form:
\begin{align} \label{eq:CR_CL_Form}
    \dot{x} &= A x + B_1 u_i + B_2 \uim,
\end{align}
where
\small{\begin{align} \label{eq:RC_CL_Matrices}
   \! \! \! \! \! A \! &= \! \! 
    \begin{bmatrix}
        0               & 1                                 & 0                                     & 0             & 0 \\ 
        0               & \frac{1}{h} - \frac{1}{\tau}      & \frac{1}{\tau} -\frac{1}{h}           & 0             & 1 \\ 
        0               & \frac{1}{h}                       & -\frac{1}{h}                          & 0             & 1 \\ 
        0               & 0                                 & 0                                     & 0             & 1 \\ 
        0               & 0                                 & 0                                     & 0             & -\frac{1}{\tau}
    \end{bmatrix}\! \!,
    \! B_1 \! = \! \!
    \begin{bmatrix}
     0                  \\ 
     -\tfrac{h}{\tau}    \\ 
     0                  \\ 
     0                  \\ 
     0                  
    \end{bmatrix}\! \!, 
    \! B_2 \! = \! \!
    \begin{bmatrix}
     0 \\ 
     0 \\ 
     0 \\ 
     0 \\ 
     \frac{1}{\tau}
    \end{bmatrix}\! \!,
\end{align}} \normalsize{}
\hspace{-8mm} and by substituting the base controller from \eqref{eq:SD_BaseController} we obtain:
\begin{align} \label{eq:RC_CL_Base}
    \begin{bmatrix}
        \dot{x} \\ \dot{\rho}_i
    \end{bmatrix} &= 
    \underbrace{\begin{bmatrix}
        A & B_1 \\
        K & -\tfrac{1}{h}
    \end{bmatrix}}_{\mathcal{A}} 
    \begin{bmatrix}
        x \\ \rho_i
    \end{bmatrix} + 
    \underbrace{\begin{bmatrix}
        B_2 \\ \tfrac{1}{h}
    \end{bmatrix}}_{\mathcal{B}_{\uim}} \uim, 
\end{align} 
as the closed-loop dynamics, where $K = [\tfrac{k_p}{h} \ \ \tfrac{k_d}{h} \ \ 0 \ \ 0 \ \ 0]$.

The sensor measurements in \eqref{eq:SD_Sensors} can be expressed in terms of the closed-loop coordinates and disturbance $\uim$, as the coordinate transformation from $[ d_i \ v_i \ a_i \ \vim \ \aim]^\top$ to $[ e_i \ \dot{e}_i \ z_i \ \vim \ \aim]^\top$ is invertible. For $\delta_{i}=0$, we obtain:
\begin{align} \label{eq:CR_SensorError}
    y_i &= 
    \underbrace{\begin{bmatrix}
        1 & 0 & -h& h & 0 \\ 
        0 & 0 & -1 & 1 & 0 \\ 
        0 & -\tfrac{1}{h} & \tfrac{1}{h} & 0 & 0 \\ 
        0 & 0 & 1 & 0 & 0 \\ 
        0 & 0 & 0 & 0 & 1 \\ 
        0 & 0 & 0 & 0 & 0
    \end{bmatrix}}_{C}
    \begin{bmatrix}
        e_i \\ \dot{e}_i \\ z_i \\ \vim \\ \aim
    \end{bmatrix} + 
    \underbrace{\begin{bmatrix}
        0 \\ 0 \\ 0 \\ 0 \\ 0 \\ 1
    \end{bmatrix}}_{D} \uim.
\end{align}
Given that $(C, \ D )$ is full rank, \eqref{eq:CR_SensorError} can be rewritten as:
\begin{align} \label{eq:InverseSensor}
    \begin{bmatrix}
        x \\ \uim
    \end{bmatrix} &= [C \ \ D ]^{-1} y_i.
\end{align}

To derive the class of base controller realizations,  $\dot{\rho}_i$ in \eqref{eq:SD_BaseController} is first reformulated using the proposed change of coordinates in \eqref{eq:CR_Transformation} and subtituting \eqref{eq:InverseSensor}, resulting in
\begin{align} \label{eq:RC_drho} \begin{split}
    \dot{\rho}_i &= -\tfrac{1}{\alpha_i h}\bar{\rho}_i + \left([ K \ \ \tfrac{1}{h} ] [C \ \ D ]^{-1} + \tfrac{1}{\alpha_i h} \beta_i\right) y_i.
\end{split} \end{align}
The dynamics of $\bar{\rho}_i$ are then obtained by differentiating \eqref{eq:CR_Transformation},
\begin{align} \begin{split}
    \dot{\bar{\rho}}_i &= \alpha_i \dot{\rho}_i + \beta_i \dot{y}_i \\
                     &= \alpha_i \dot{\rho}_i + \beta_i (C \dot{x} + D \dot{u}\imo),
\end{split}\end{align}
where due to the structure of $\beta_i$ and $D$ we have that $\beta_i D = 0$. Substitution of \eqref{eq:CR_CL_Form} and \eqref{eq:InverseSensor} provides that:
\begin{align}\label{eq:RC_drho_bar1} \begin{split}
    \dot{\bar{\rho}}_i &= \alpha_i \dot{\rho}_i + \beta_i C [ A \ \ B_2 ] [C \ \ D ]^{-1} y_i  + \beta_i C B_1 u_i.
\end{split}\end{align}
The new formulation of the control input $u_i$ is obtained by applying the change of coordinates to \eqref{eq:SD_BaseController}, resulting in
\begin{subequations} \label{eq:RealizationClass}   
\begin{align} \label{eq:RC_ui}
    u_i \! &= \! \tfrac{1}{\alpha_i} \bar{\rho}_i - \tfrac{1}{\alpha_i} \beta_i y_i. 
\intertext{Substitution of \eqref{eq:RC_drho} and \eqref{eq:RC_ui} into \eqref{eq:RC_drho_bar1} yields:}
\begin{split} \label{eq:RC_drho_bar2} 
     \dot{\bar{\rho}}_i &= (\tfrac{1}{\alpha_i} \beta_i C B_1  -\tfrac{1}{h}) \bar{\rho}_i + \left( \left( \alpha_i [ K \ \ \tfrac{1}{h} ] + \right. \right. \\ 
    & \left. \left. \beta_i C [ A \ \ B_2 ] \right) [ C \ \ D ]^{-1}  
    \! \! + \! \tfrac{1}{h} \beta_i \! - \! \tfrac{1}{\alpha_i} \beta_i C B_1 \beta_i \right) y_i.
\end{split}\end{align}
\end{subequations}
Consequently, \eqref{eq:RealizationClass} represents the class of controller realizations of \eqref{eq:SD_BaseController}, which exhibit equivalent closed-loop behavior without an FDI attack. Notice that for $\alpha_i = 1, \beta_i = 0$, or for $\alpha_i = -\tfrac{\tau}{h}, \beta_i = [0 \ \ 0 \ \ (1-\tfrac{\tau}{h}) \ \ 0 \ \ \tfrac{\tau}{h} \ \ 0]$, the controller realizations $\mathcal{C}$ or $\hat{\mathcal{C}}$ are obtained, respectively. Note that, \eqref{eq:RealizationClass} is derived for $\delta_i = 0$ to find all equivalent controller realizations that exhibit equivalent closed-loop behavior. 

However, the change of coordinates affects the combinations of sensors being used and, therefore, also influences the effect of an FDI on the closed-loop behavior. To model the effect of FDI attacks, we now derive \eqref{eq:RealizationClass} for $\delta_i \neq 0$ by reformulating \eqref{eq:CR_SensorError} by including FDI attack signals $\delta_i$:
\begin{align} \label{eq:CR_y_att}
    y_i &= [C \ \ D ] \begin{bmatrix}
        x \\ \uim
    \end{bmatrix} + \delta_i.
\end{align}
Substitution of \eqref{eq:CR_y_att} into \eqref{eq:RealizationClass} yields the realized controller in presence of FDI attacks, 
\begin{subequations} \label{eq:RealizedController} \begin{align}
    u_i \! &= \! \tfrac{1}{\alpha_i} \bar{\rho}_i - \tfrac{1}{\alpha_i} \beta_i [C \ \ D ] \begin{bmatrix} x \\ \uim \end{bmatrix} - \tfrac{1}{\alpha_i} \beta_i \delta_i, \\
    \begin{split} 
        \dot{\bar{\rho}}_i \! &= \! (\tfrac{1}{\alpha_i} \beta_i C B_1  -\tfrac{1}{h}) \bar{\rho}_i + \left( \alpha_i [ K \ \ \tfrac{1}{h} ] + \beta_i C [ A \ \ B_2 ] \right. \\
        & \ \ \left. + (\tfrac{1}{h} \beta_i -\tfrac{1}{\alpha_i} \beta_i C B_1 \beta_i)[C \ \ D ] \right) \begin{bmatrix} x \\ \uim \end{bmatrix} \\
        & \ \ + \left( \left( \alpha_i [ K \ \ \tfrac{1}{h} ] + \beta_i C [ A \ \ B_2 ] \right) [C \ \ D ]^{-1}  \right. \\
        & \ \ \left. + \tfrac{1}{h} \beta_i - \tfrac{1}{\alpha_i} \beta_i C B_1 \beta_i \right) \delta_i.
\end{split}\end{align}\end{subequations}

When deriving the closed-loop system using \eqref{eq:CR_CL_Form} and \eqref{eq:RealizedController}, the system matrices $\bar{\mathcal{A}}(\alpha_i, \beta_i)$ and $\bar{\mathcal{B}}_{\uim}(\alpha_i, \beta_i)$ are dependent on the choice of $\alpha_i$ and $\beta_i$. Although different sensors are used, the realization does not affect the nominal closed-loop behavior. This is because the input-output behavior of the closed-loop system is invariant under a linear coordinate transformation, commonly referred to as a similarity transformation. Therefore, after selecting the required combination of sensors (including $\delta_i$), the closed-loop system is transformed back to its original coordinates $[x \ \ \rho_i]^\top$ as in \eqref{eq:RC_CL_Base}. As we show below: (i) a fair comparative analysis is made between different realizations, as $\mathcal{{A}}$ and $\mathcal{B}_{\uim}$ in \eqref{eq:RC_CL_Base} are independent of $\alpha_i$ and $\beta_i$, and therefore only $\mathcal{B}_{\delta_i}$ is dependent on $\alpha_i$ and $\beta_i$, and (ii) the resulting dynamics are affine in $\tfrac{1}{\alpha_i}\beta_i$, allowing for linear and convex optimization techniques, whereas solving it in the new coordinates requires nonlinear optimization techniques. To prove the latter, note that the change of coordinates in \eqref{eq:CR_Transformation} is invertible for $\alpha_i \neq 0$:
\begin{align}
\begin{bmatrix}
        x \\ \bar{\rho}_i
    \end{bmatrix}&=
    \begin{bmatrix}
        I & 0 \\
        \beta_iC & \alpha_i
    \end{bmatrix} \begin{bmatrix}
        x \\ \rho_i
    \end{bmatrix}. 
\end{align}
By applying the inverse coordinate transformation to the closed-loop system of \eqref{eq:CR_CL_Form} using \eqref{eq:RealizedController}, the original closed-loop system \eqref{eq:RC_CL_Base} is obtained. However, now the closed-loop dynamics include a matrix that represents the effect of an FDI attack on the closed-loop system dynamics (depending on the realization of the controller):
\begin{subequations} 
   \begin{align} \label{eq:RC_CL_Att} 
    & \begin{bmatrix}
        \dot{x} \\ \dot{\rho}_i
    \end{bmatrix} = 
    \mathcal{A} 
    \begin{bmatrix}
        x \\ \rho_i
    \end{bmatrix} + 
    \mathcal{B}_{\uim} \uim + \mathcal{B}_{\delta_i} \delta_i, \\%
& \! \! \! \! \text{where}  \nonumber \\
 \label{eq:RC_Bdelta}
    & \! \!  \! \! \mathcal{B}_{\delta_i} \! = \! \! 
    \begin{bmatrix}
        -\tfrac{1}{\alpha_i} B_1 \beta_i \\
        \left([ K \ \ \tfrac{1}{h} ] \! + \! \tfrac{1}{\alpha_i} \beta_i C [ A \ \ B_2 ]\right) \! [C \ \ D ]^{-1} \! \! + \tfrac{1}{h \alpha_i} \beta_i
    \end{bmatrix} \! \!.
\end{align} 
\end{subequations}

Note that when converting the system back to its original coordinates, only the attack matrix $\mathcal{B}_{\delta_i}$ is affected by the choice of $\alpha_i$ and $\beta_i$. The following section introduces an optimization scheme to find the optimal realization of the base controller within the suggested class of realized controllers. 
\vspace{-2mm}

\subsection{Optimization Problem} \vspace{-1mm}
When looking for the optimal controller realization, we aim to minimize the reachable set induced by resource-limited FDI attacks, \textcolor{red}{which manipulate sensing, actuation, and networked data while being constrained by physical limitations, computing power, and attack strategy \cite{zhang_networked_2019}}. We use the volume of these reachable sets as a security metric, whereas minimizing the volume directly minimizes the size of the state space portion that can be induced by a series of attacks \cite{murguia_security_2020}. Computing the exact reachable set is generally not tractable and time-dependent. Instead, we seek to minimize the volume of the outer ellipsoidal approximation.

Computing the reachable set for \eqref{eq:RC_CL_Att} is not possible, due to $\vim$ and $\aim$ being uncontrollable states, introducing zero-eigenvalues in $\mathcal{A}$. However, note that $\vim$ and $\aim$ are fully decoupled (see \eqref{eq:RC_CL_Base}). Moreover, the choice of $\alpha_i$ and $\beta_i$ does not affect $\vim$ and $\aim$, as the attack only affects the dynamics of $\dot{e}_i$ and $\rho_i$. Therefore, \eqref{eq:RC_CL_Att} is reformulated
\begin{align} \label{eq:OR_DecoupledDynamics}
        \dot{x}_i  &= \mathcal{A}_i x_i + \mathcal{B}_{i-1} 
    \begin{bmatrix}
        x\imo \\ \uim
    \end{bmatrix} + B_{i,\delta_i} \delta_i, 
\end{align}
where $x_i = [e_i \ \dot{e}_i \ z_i \ \rho_i]^\top$, $x\imo = [\vim \ \aim]^\top$, and $\mathcal{B}_{i,\delta_i}$ as in \eqref{eq:RC_Bdelta} but excluding the zero entries for $\dvim$ and $\daim$. 

In \eqref{eq:OR_DecoupledDynamics}, $x\imo$ and $\uim$ act as input on the platooning dynamics. Assuming boundedness of $x\imo$ and $\uim$, which is reasonable considering vehicle physical constraints, a reachable set can be found for $\delta_i = 0$ due to the additional disturbances defined by $\mathcal{B}_{\imo}$. However, the ellipsoidal approximation of the reachable set for \eqref{eq:OR_DecoupledDynamics} with $\delta_i \! \neq \! 0$ can be obtained by the Minkowski sum of two independent ellipsoidal approximation of the reachable induced by $\mathcal{B}_{\imo}$ and $\mathcal{B}_{\delta_i}$ separately \cite{halder_parameterized_2018}. Therefore, we mitigate the effect of $\mathcal{B}_{\imo}$, as the optimal $\alpha_i$ and $\beta_i$ do not affect this reachable set. Moreover, we are only interested in minimizing the reachable set induced by $\mathcal{B}_{\delta_i}$. To this end, for synthesis of the optimal realization, \eqref{eq:OR_DecoupledDynamics} is modeled as it is only perturbed by $\delta_i$
\begin{align}\label{eq:CL_ReachableSet1}
    \dot{\tilde{x}}_i &= \mathcal{A}_i \tilde{x}_i + \mathcal{B}_{i,\delta_i} \delta_{i},
\end{align}
where $\tilde{x}_i$ is used to distinguish the platooning dynamics in \eqref{eq:OR_DecoupledDynamics} and \eqref{eq:CL_ReachableSet1}, as these represent different dynamical systems.

Given that the realized controller operates in discrete time, and attacks operate on sampled signals, a discrete-time equivalent model of \eqref{eq:CL_ReachableSet1} is obtained via exact discretization at the sampling time instant, $t = T_s k, \, k \in \mathbb{N}$, where $T_s>0$ is the sampling interval. The resulting discrete-time model can be written as
\begin{align} \label{eq:CL_ReachableSet2}
    \tilde{x}_i(k+1)  &= \mathcal{A}_i^d  \tilde{x}_i(k) + \mathcal{B}^d_{i,\delta_i} \delta_{i}(k), \ \tilde{x}_i(0) = \tilde{x}_{i,0}
\end{align}
with
\begin{equation}
\begin{aligned}
\label{eq:DiscreteSystemMatrices}
& \mathcal{A}_i^d=e^{\mathcal{A}_i T_s}, 
& \mathcal{B}^d_{i,\delta_i} = \left(\int_0^{T_s} e^{\mathcal{A}_i\left(T_s-s\right)} ds \right) \mathcal{B}_{i,\delta_i}.
\end{aligned}
\end{equation}

The constraints we impose on $\delta_i(k)$ have the following structure
\begin{align} \label{eq:DeltaBound} 
        & \delta_{i,j}(k) \in \{ \delta_{i,j} \mid \hspace{1mm}  \delta^2_{i,j} \leq W^2_{i,j} \}, \, \forall k \in \mathbb{N},  
\end{align}
for some known constants $W_{i,j}\in\mathbb{R}_{>0}, j \in \{1,2,..,6\}$. Associated with these constraints, we introduce the adversarial reachable set (Definition~\ref{def1})
\begin{equation}\label{eq:reachable_set}
\mathcal{R}^{\tilde{x}_{i,0}}(k):=\left\{\tilde{x}_i \in \mathbb{R}^4 \left|
\begin{aligned}
    & \tilde{x}_{i,0} = \tilde{x}_i(k) ,  \\ 
    & \tilde{x}_i(k) \text{ solution to } \eqref{eq:CL_ReachableSet2},\\
    & \delta_{i}(k) \text{ satisfies } \eqref{eq:DeltaBound}.
\end{aligned}
\right. \right\}
\end{equation}
We seek to obtain the outer ellipsoidal approximation $\mathcal{E}^{\tilde{x}_i}(k)$ of $\mathcal{R}^{\tilde{x}_i}(k)$ via Lemma~\ref{lemma1}. Moreover, we have that  $\mathcal{E}^{\tilde{x}_i}(k) \approx \mathcal{E}^{\tilde{x}_i}(\infty) \coloneqq \{ \tilde{x}_i \mid \tilde{x}_i^\top \mathcal{P}^{\tilde{x}_i} \tilde{x}_i \leq (N-a)/(1-a) \}$, for some positive definite matrix $\mathcal{P}^{\tilde{x}_i} \in \mathbb{R}^{4 \times 4}$,\textcolor{red}{where $N$ represents the number of disturbances, and $a \in (0, 1)$.} As we consider the reachable set at infinity, the ellipsoidal approximation is independent of the initial condition.

By applying Lemma~\ref{lemma1}, we can find $\mathcal{P}^{\tilde{x}_i}$, where $\mathcal{A} = \mathcal{A}^d_{i}$, $\mathcal{B} = \mathcal{B}^d_{i,\delta_i}$, with $\mathcal{A}^d_i$ and $\mathcal{B}^d_{i,\delta_i}$ as formulated in \eqref{eq:DiscreteSystemMatrices}. However, now we aim to minimize $-\log\det[\mathcal{P}^{\tilde{x}_i}]$ while also optimizing $\mathcal{B}(\alpha_i, \beta_i)$, which results in a non-linear control problem. To this end, we convert the SDP problem in Lemma~\ref{lemma1} to a problem formulation that is affine in both $\mathcal{P}^{\tilde{x}_i}$ and $\mathcal{B}(\alpha_i, \beta_i)$. A congruence transformation of the form $Q W Q^\top \geq 0$ is applied, where $Q = Q^\top = \diag [(\mathcal{P}^{\tilde{x}_i})^{-1}, (\mathcal{P}^{\tilde{x}_i})^{-1}, I] >0$, which preserves the definiteness of the matrix inequality \cite{Boyd1994}. Applying this congruence transformation to Lemma~\ref{lemma1} yields
    \begin{align}
     \begin{bmatrix}
			a (\mathcal{P}^{\tilde{x}_i})^{-1}                       & (\mathcal{P}^{\tilde{x}_i})^{-1} (\mathcal{A}_i^d)^{\top}     & \mathbf{0}\\
			\mathcal{A}_i^d (\mathcal{P}^{\tilde{x}_i})^{-1}         & (\mathcal{P}^{\tilde{x}_i})^{-1}                              & \mathcal{B}(\alpha_i, \beta_i) \\
			\mathbf{0}                                               & (\mathcal{B}(\alpha_i, \beta_i))^{\top}                       & W_{a}
	\end{bmatrix} &\geq 0.
\end{align}
However, when applying the congruence transformation, the resulting objective function $\min -\log\det[(\mathcal{P}^{\tilde{x}_i})^{-1}]$ becomes concave. Instead, we seek to minimize a convex upper bound on $\sqrt{\det[(\mathcal{P}^{\tilde{x}_i})^{-1}]}$, which is proportional to the volume of the ellipsoid, from which also the original objective $\log \det [\mathcal{P}^{\tilde{x}_i}]$ is derived. To this end, we use Lemma~\ref{lemma3} and minimize $\trace[Y]$, where $Y = (\mathcal{P}^{\tilde{x}_i})^{-1}$. Additionally, $\mathcal{B}(\alpha_i, \beta_i)$ is affine in $\tfrac{1}{\alpha_i}\beta_i$, where $\alpha_i$ only scales $\beta_i$ and hence does not affect the reachable set. Therefore, w.l.o.g. we solve the SDP for $\alpha_i = 1$. As a result, we obtain 
\begin{align} \label{eq:OP_OptProblem}
	\left\{\!\begin{aligned}
        &\underset{Y, \beta_i,  a_{1},\ldots, a_{N}}{\text{minimize}} \trace[Y],\\
		&\text{\emph{s.t.}} \hspace{1mm}a_{1},\ldots, a_{N}\in(0,1), \hspace{.5mm} a_{1}+\ldots+ a_{N}\geq a,\\
		&Y>0, 
     \begin{bmatrix}
			a Y                       & Y (\mathcal{A}_i^d)^{\top}               & \mathbf{0}\\
			\mathcal{A}_i^d Y         & Y                                        & \mathcal{B}(\beta_i) \\
			\mathbf{0}                & (\mathcal{B}(\beta_i))^{\top}      & W_{a}
	\end{bmatrix} \geq 0.
	\end{aligned}\right.
\end{align}

This minimization is solved in the following section to optimize the realization of $\mathcal{C}$.

}
{

\subsection{Optimization Problem} 
In search of the optimal controller realization, we aim to minimize the reachable set induced by resource-limited FDI attacks on the closed-loop dynamics. FDI attacks manipulate sensing, actuation, and networked data while constrained by physical limitations, computing power, and attack strategy \cite{zhang_networked_2019}. We use the volume of these reachable sets as a security metric, whereas minimizing the volume directly reduces the size of the state space portion that can be induced by a series of attacks \cite{murguia_security_2020}. Computing the exact reachable set is generally not tractable and time-dependent. Instead, we seek to minimize the volume of the outer ellipsoidal approximation of the reachable set.

Computing the reachable set for \eqref{eq:RC_CL_Att} is not possible, due to $\vim$ and $\aim$ being uncontrollable states, introducing zero-eigenvalues in $\mathcal{A}$. However, note that $\vim$ and $\aim$ are fully decoupled from the other states (see \eqref{eq:RC_CL_Matrices}, \eqref{eq:RC_CL_Base}). Moreover, the choice of $\alpha_i$ and $\beta_i$ does not affect $\vim$ and $\aim$, as the attack only affects the dynamics of $\dot{e}_i$ and $\rho_i$. Therefore, \eqref{eq:RC_CL_Att} is reformulated
\begin{align} \vspace{-2mm} \label{eq:OR_DecoupledDynamics}
        \begin{bmatrix} \dot{x}_i \\ \dot{\rho}_i \end{bmatrix}  &= \mathcal{A}_i \begin{bmatrix} x_i \\ \rho_i \end{bmatrix} + \mathcal{B}_{i,i-1} 
    \begin{bmatrix}
        x\imo \\ \uim
    \end{bmatrix} + \mathcal{B}_{i,\delta_i} \delta_i, 
\vspace{-2mm}  \end{align}
where $x_i = [e_i \ \dot{e}_i \ z_i]^\top$, $x\imo = [\vim \ \aim]^\top$, $\mathcal{A}_i$ and $\mathcal{B}_{i,i-1}$ derived from  \eqref{eq:RC_CL_Base}, and $\mathcal{B}_{i,\delta_i}$ as $\mathcal{B}_{\delta_i}$ in \eqref{eq:RC_Bdelta} but excluding the zero entries for $\dvim$ and $\daim$. 

In \eqref{eq:OR_DecoupledDynamics}, $x\imo$ and $\uim$ act as input on the platooning dynamics. Assuming boundedness of $x\imo$ and $\uim$, which is reasonable considering vehicle physical constraints, a reachable set can be found for $\delta_i = 0$ due to the additional disturbances entering through $\mathcal{B}_{i,i-1}$. However, the ellipsoidal approximation of the reachable set for \eqref{eq:OR_DecoupledDynamics} with $\delta_i \! \neq \! 0$ can be obtained by the Minkowski sum of two independent ellipsoidal approximations of the reachable sets induced by the inputs associated to $\mathcal{B}_{i,i-1}$ and $\mathcal{B}_{i,\delta_i}$ separately \cite{halder_parameterized_2018}. Therefore, we discard the effect of $\mathcal{B}_{i,i-1}$, as the optimal $\alpha_i$ and $\beta_i$ do not affect this reachable set. Moreover, we are only interested in minimizing the reachable set induced by $\mathcal{B}_{i,\delta_i}$ to enhance robustness against FDI attacks. To this end, for the synthesis of the optimal realization, \eqref{eq:OR_DecoupledDynamics} is modeled as if it is only perturbed by $\delta_i$:
\begin{align} \vspace{-2mm} \label{eq:CL_ReachableSet1}
    \begin{bmatrix} \dot{\tilde{x}}_i \\ \dot{\tilde{\rho}}_i\end{bmatrix} &= \mathcal{A}_i     \begin{bmatrix} \tilde{x}_i \\ \tilde{\rho}_i\end{bmatrix}  + \mathcal{B}_{i,\delta_i} \delta_{i},
\vspace{-2mm} \end{align}
where $\tilde{x}_i$ is used to distinguish the platooning dynamics in \eqref{eq:OR_DecoupledDynamics} and \eqref{eq:CL_ReachableSet1}, as these represent different dynamical systems.

Given the fact that the realized controller operates in discrete time, and attacks operate on sampled signals, a discrete-time equivalent model of \eqref{eq:CL_ReachableSet1} is obtained via exact discretization at the sampling time instant, $t = T_s k, \, k \in \mathbb{N}$, where $T_s>0$ is the sampling interval. We assume a zero-order hold on control input $u(t)$ and sampling on the FDI attack $\delta_i(t)$. The resulting discrete-time model yields:
\begin{subequations}
\begin{align} \vspace{-2mm} \label{eq:CL_ReachableSet2}
    \tilde{x}_i(k+1)  &= \mathcal{A}_i^d  \tilde{x}_i(k) + \mathcal{B}^d_{i,\delta_i} \delta_{i}(k), \ \tilde{x}_i(0) = \tilde{x}_{i,0}
\vspace{-2mm} \end{align}
with
\begin{equation}
\begin{aligned}
\label{eq:DiscreteSystemMatrices}
& \mathcal{A}_i^d=e^{\mathcal{A}_i T_s}, 
& \mathcal{B}^d_{i,\delta_i} = \left(\int_0^{T_s} e^{\mathcal{A}_i\left(T_s-s\right)} ds \right) \mathcal{B}_{i,\delta_i}.
\end{aligned}
\end{equation}
\end{subequations}

On $\delta_i(k)$ we impose the constraints of the form
\begin{align} \vspace{-1mm} \label{eq:DeltaBound} 
        & \delta_{i,j}(k) \in \{ \delta_{i,j} \mid \hspace{1mm}  \delta^2_{i,j} \leq W^2_{i,j} \}, \, \forall k \in \mathbb{N},  
\vspace{-1mm} \end{align}
for some known constants $W_{i,j}\in\mathbb{R}_{>0}, j \in \{1,2,..,6\}$. Associated with these constraints, the adversarial reachable set (Definition~\ref{def1}) is introduced
\begin{equation}\label{eq:reachable_set}
\mathcal{R}^{\tilde{x}_{i,0}}(k):=\left\{\tilde{x}_i \in \mathbb{R}^4 \left|
\begin{aligned}
    & \tilde{x}_{i,0} = \tilde{x}_i(0) ,  \\ 
    & \tilde{x}_i(k) \text{ solution to } \eqref{eq:CL_ReachableSet2},\\
    & \delta_{i}(k) \text{ satisfies } \eqref{eq:DeltaBound}.
\end{aligned}
\right. \right\}
\end{equation}
We seek to obtain the outer ellipsoidal approximation $\mathcal{E}^{\tilde{x}_i}(k)$ of $\mathcal{R}^{\tilde{x}_{i,0}}(k)$ via Lemma~\ref{lemma1}. Moreover, we have that  $\mathcal{E}^{\tilde{x}_i}(k) \approx \mathcal{E}^{\tilde{x}_i}(\infty) \coloneqq \{ \tilde{x}_i \mid \tilde{x}_i^\top \mathcal{P}^{\tilde{x}_i} \tilde{x}_i \leq (N-a)/(1-a) \}$, for some positive definite matrix $\mathcal{P}^{\tilde{x}_i} \in \mathbb{R}^{4 \times 4}$. Considering the reachable set at infinity, the ellipsoidal approximation is independent of the initial condition, given that $\mathcal{A}_i^d$ is Schur.

By applying Lemma~\ref{lemma1}, $\mathcal{P}^{\tilde{x}_i}$ can be found, where $\mathcal{A} = \mathcal{A}^d_{i}$, $\mathcal{B} = \mathcal{B}^d_{i,\delta_i}$, with $\mathcal{A}^d_i$ and $\mathcal{B}^d_{i,\delta_i}$ as formulated in \eqref{eq:DiscreteSystemMatrices}. However, the goal is to solve \eqref{lmi1} by minimizing $-\log\det[\mathcal{P}^{\tilde{x}_i}]$, while also optimizing $\alpha_i$ and  $\beta_i$, leading to a non-linear optimization problem. To this end, the SDP problem in Lemma~\ref{lemma1} is converted to a problem formulation that is affine in both $\mathcal{P}^{\tilde{x}_i}$ and $\mathcal{B}^d_{i,\delta_i}(\alpha_i, \beta_i)$. A congruence transformation of the form $Q W Q^\top \geq 0$ is applied, with $W \geq 0$ and $Q = Q^\top = \diag [(\mathcal{P}^{\tilde{x}_i})^{-1}, (\mathcal{P}^{\tilde{x}_i})^{-1}, I] >0$, which preserves the definiteness of the matrix inequality \cite{Boyd1994}. Applying this congruence transformation to Lemma~\ref{lemma1} yields
    \begin{align} \vspace{-2mm}
     \begin{bmatrix}
			a (\mathcal{P}^{\tilde{x}_i})^{-1}                       & (\mathcal{P}^{\tilde{x}_i})^{-1} (\mathcal{A}_i^d)^{\top}     & \mathbf{0}\\
			\mathcal{A}_i^d (\mathcal{P}^{\tilde{x}_i})^{-1}         & (\mathcal{P}^{\tilde{x}_i})^{-1}                              & \mathcal{B}^d_{i,\delta_i}(\alpha_i, \beta_i) \\
			\mathbf{0}                                               & (\mathcal{B}^d_{i,\delta_i}(\alpha_i, \beta_i))^{\top}                       & W_{a}
	\end{bmatrix} &\geq 0,
\vspace{-2mm} \end{align}
which is now linear in both $\mathcal{P}^{\tilde{x}_i}$ and $\mathcal{B}^d_{i,\delta_i}$. Moreover, the resulting SDP is now an LMI.

However, when applying the congruence transformation, the resulting objective function $\min -\log\det[(\mathcal{P}^{\tilde{x}_i})^{-1}]$ becomes concave. Therefore, we alternatively seek to minimize a convex upper bound on $\sqrt{\det[(\mathcal{P}^{\tilde{x}_i})^{-1}]}$, which is proportional to the volume of the ellipsoid, from which also the original objective $\log \det [\mathcal{P}^{\tilde{x}_i}]$ is derived. To this end, we minimize $\trace[Y]$, where $Y = (\mathcal{P}^{\tilde{x}_i})^{-1}$ \cite{murguia_security_2020}. Additionally, $\mathcal{B}^d_{i,\delta_i}(\alpha_i, \beta_i)$ is affine in $\tfrac{1}{\alpha_i}\beta_i$, where $\alpha_i$ only scales $\beta_i$ and hence does not affect the reachable set. Therefore, w.l.o.g., the SDP is solved for $\alpha_i = 1 \ \forall i \in S_m \backslash \{1\}$, resulting in the following optimization problem:
\begin{align} \label{eq:OP_OptProblem}
	\left\{\!\begin{aligned}
        &\underset{Y, \beta_i,  a_{1},\ldots, a_{N}}{\text{minimize}} \trace[Y],\\
		&\text{\emph{s.t.}} \hspace{1mm}a_{1},\ldots, a_{N}\in(0,1), \hspace{.5mm} a_{1}+\ldots+ a_{N}\geq a,\\
		&Y>0, 
     \begin{bmatrix}
			a Y                       & Y (\mathcal{A}_i^d)^{\top}               & \mathbf{0}\\
			\mathcal{A}_i^d Y         & Y                                        & \mathcal{B}^d_{i,\delta_i} (\beta_i) \\
			\mathbf{0}                & (\mathcal{B}^d_{i,\delta_i}(\beta_i))^{\top}      & W_{a}
	\end{bmatrix} \geq 0.
	\end{aligned}\right.
\end{align}

This minimization problem is solved in the following section to optimize the realization of $\mathcal{C}$.

}

\section{Results} \label{sec:Results}
\runScript{

In this section, we solve the optimization problem introduced in $\eqref{eq:OP_OptProblem}$ to determine the optimal realization of \eqref{eq:SD_BaseController}. We adopt the controller settings from \cite{ploeg_design_2011}, with a desired inter-vehicle distance of $r = 3\,$m, with driveline dynamics constant $\tau = 0.1\,$s, time headway constant $h = 0.5\,$s, and controller gains of $(k_p, k_d) = (0.2, 0.7)$. The sampling rate is $T_s = 0.01\,$s. For the resource-limited FDI attacks $\delta_{i,j}(k)$, we assume adherence to the bound specified in \eqref{eq:DeltaBound} with $W_{i} = I$, thereby considering FDI attacks on all sensors. The bound $W_{i,j}$ is arbitrarily chosen to illustrate the results. 

The results are obtained using YALMIP \cite{YALMIP}, with SDP solver SDPT3 \cite{SDPT3}. Setting $\alpha_i = 1$, we obtain the following optimal realization
\begin{align} \label{eq:R_Optimal}
   \! \! \bar{\mathcal{C}} \! &\coloneqq \! \left\{ 
   \begin{aligned}
        \bar{u}_i &= \! \bar{\rho}_i \! + \!  0.771 y_{i,1} \!  - \!  0.33 y_{i,2} \\ 
        &  -0.135 y_{i,3} + 1.672 y_{i,4} + 0.187 y_{i,5}, \\
        \dot{\bar{\rho}}_i &= -0.65\bar{\rho}_i - 1.142 y_{i,1} + 0.46 y_{i,2} \\
         & \! + \! 0.222 y_{i,3} \! - \! 2.715 y_{i,4} \! - \! 0.176 y_{i,5} \! + \! 0.13 y_{i,6},
    \end{aligned} \right.
\end{align}
Notably, the optimal realization $\bar{\mathcal{C}}$ maximizes robustness against $\delta_i$ by utilizing all available sensors \eqref{eq:CR_SensorError}. 

In Fig.~\ref{fig:ReachableSet} the outer ellipsoidal approximation of \eqref{eq:CL_ReachableSet2} (using Lemma~\ref{lemma1}) is projected on the $e_i$-$z_i$ plane (using Lemma~\ref{lemma2}). Among the three controllers, despite being derived from minimizing the upper bound on the volume, $\bar{\mathcal{C}}$ has the smallest ellipsoidal approximation $(\bar{\mathcal{E}}^{\tilde{x}_i})$ in terms of volume, therefore being the most robust realization. However, the projections in Fig.~\ref{fig:ReachableSet} indicate there exists some attack where $\mathcal{C}$ is more robust than $\bar{\mathcal{C}}$, as $\bar{\mathcal{E}}^{\tilde{x}_i} \nsubseteq \mathcal{E}^{\tilde{x}_i}$.
\begin{figure}[bt]\centering
		\includegraphics[width=\linewidth]{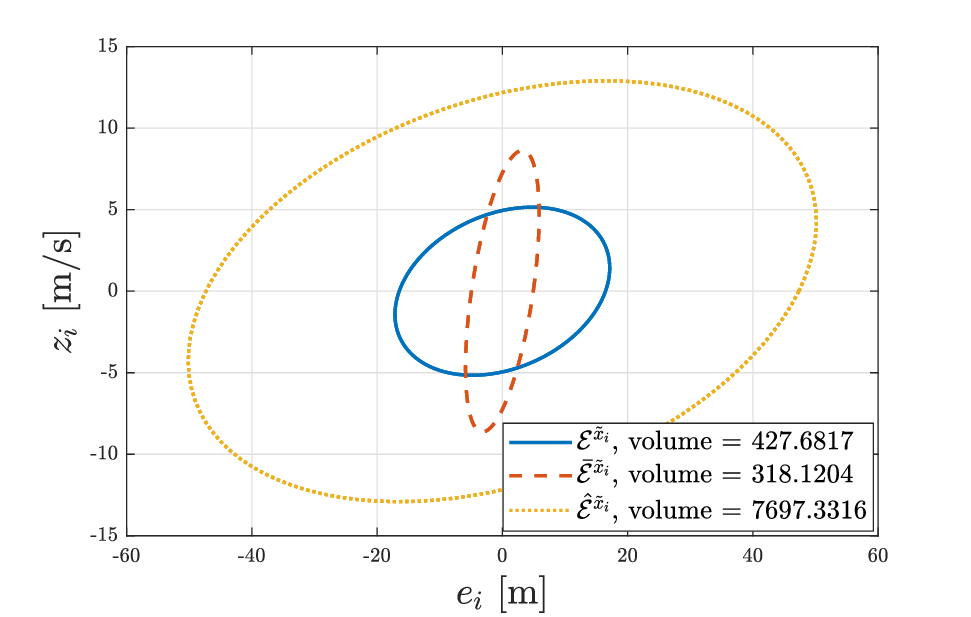}
		\caption{Outer ellipsoidal approximation of the reachable set in \eqref{eq:CL_ReachableSet2} for the different controller realizations $\mathcal{C}$, $\bar{\mathcal{C}}$, and $\hat{\mathcal{C}}$. Results are projected onto the $z_i$-$e_i$ plane, given that $\delta_i \neq 0$, and $W_i = I$.}
		\centering
		\label{fig:ReachableSet}
\end{figure}

As a case study, all three controllers are simulated using MATLAB Simulink. The vehicles are modeled via the platooning dynamics in \eqref{eq:SD_PlatoonDynamics}, where the controllers in  \eqref{eq:SD_BaseController}, \eqref{eq:C2_Lefeber}, and \eqref{eq:R_Optimal} are used to control three different platoons. Each platoon consists of a follower (vehicle $i$), controlled via one of the realized controllers, and a leader vehicle (vehicle $i-1$), controlled via a cruise controller. The leader aims to drive a steady-state velocity of 50 [km/h] while being subject to some traffic in the initial stage, resulting in some transient platooning behavior. Two different scenarios are investigated: (i) vehicle $i$ is subject to a FDI attack on $y_{i,2}$, representing the onboard velocity measurement, where $\delta_{i,2} = \sin(0.01 t) \ \forall t \in [20, 150]$, and (ii) vehicle $i$ is subject to a FDI attack on $y_{i,3}$, representing the onboard acceleration measurement, where $\delta_{i,3} = \sin(3 t)\ \forall t\in [20, 75]$. 

The results for scenario (i) and scenario (ii) are depicted in Fig.~\ref{fig:Att_y2} and Fig.~\ref{fig:Att_y3}, respectively. The results show at the initial stage that the different controller realizations exhibit equivalent platooning behavior, as the platooning behavior is invariant under the coordinate transformation. During the FDI in scenario (i), the vehicle controlled by $\bar{\mathcal{C}}$ is the most robust against the attack, as the vehicle states and control input in Fig.~\ref{fig:Att_y2} show the least magnitude amplification of $\delta_{i,2}$. In scenario (ii), controller $\mathcal{C}$ shows the least magnitude amplification of $\delta_{i,3}$, which corresponds with the results obtained in Fig.~\ref{fig:ReachableSet}, where $\bar{\mathcal{E}}^{\tilde{x}_i} \nsubseteq \mathcal{E}^{\tilde{x}_i}$. 

\begin{figure}[bt]\centering
		\includegraphics[width=\linewidth]{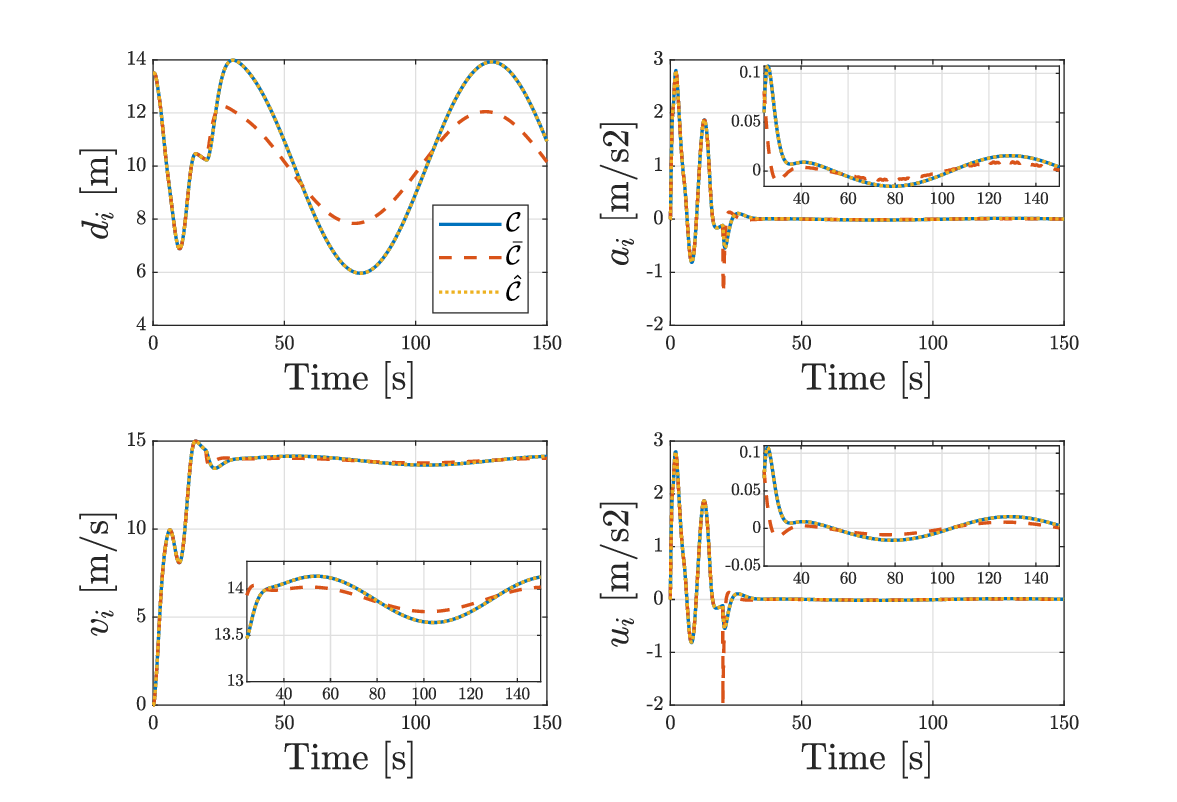}
		\caption{Vehicle and controller response for the different controller realizations $\mathcal{C}$, $\bar{\mathcal{C}}$, and $\hat{\mathcal{C}}$, given an FDI attack on $y_{i,2}$, where $\delta_{i,2} = \sin(0.01 t) \ \forall t \in [20, 150]$ s.}
		\centering
		\label{fig:Att_y2}
\end{figure}
\begin{figure}[bt]\centering
		\includegraphics[width=\linewidth]{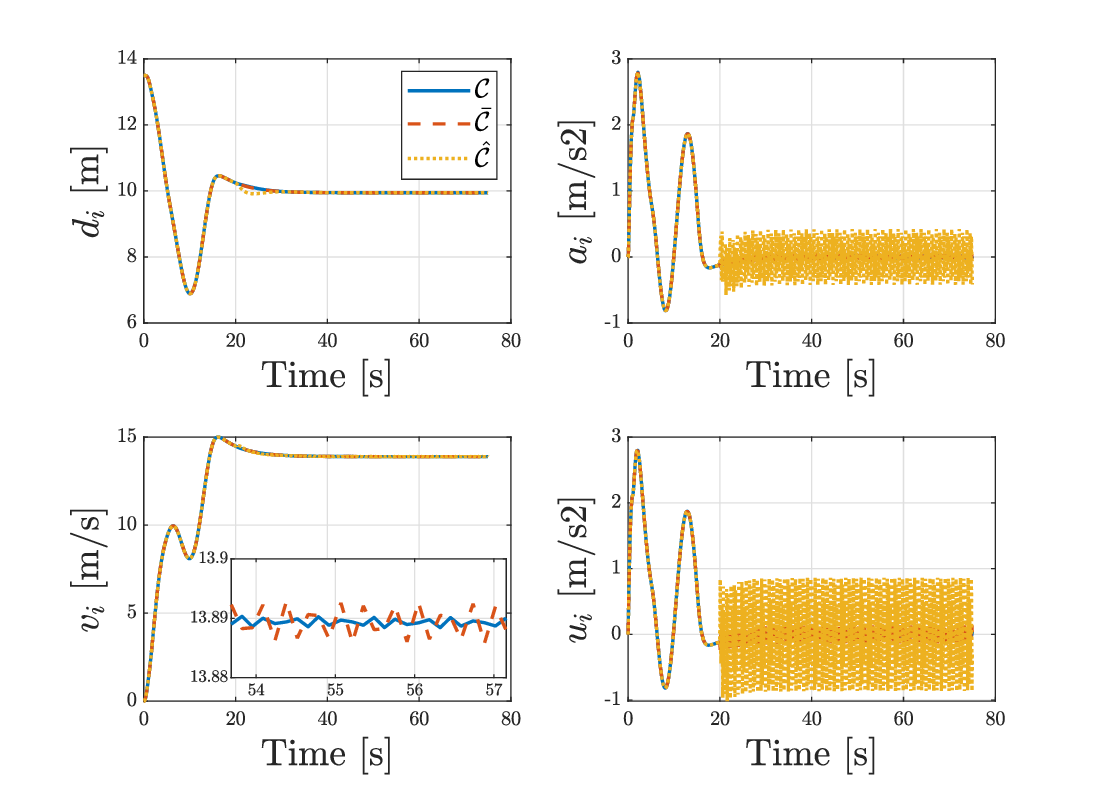}
		\caption{Vehicle and controller response for the different controller realizations $\mathcal{C}$, $\bar{\mathcal{C}}$, and $\hat{\mathcal{C}}$, given an FDI attack on $y_{i,3}$, where $\delta_{i,3} = \sin(3 t) \ \forall t \in [20, 75]$ s.}
		\centering
		\label{fig:Att_y3}
\end{figure}

}{
In this section, the optimization problem introduced in $\eqref{eq:OP_OptProblem}$ is solved to determine the optimal realization of the base controller in \eqref{eq:SD_BaseController}. We adopt the controller settings from \cite{ploeg_design_2011}, with a desired inter-vehicle distance of $r = 3\,$m, with driveline dynamics constant $\tau = 0.1\,$s, time headway constant $h = 0.5\,$s, and controller gains of $(k_p, k_d) = (0.2, 0.7)$. The sampling rate is $T_s = 0.01\,$s. For the resource-limited FDI attacks $\delta_{i,j}(k)$, we assume adherence to the bound specified in \eqref{eq:DeltaBound} with $W_{i} = I$, thereby considering FDI attacks on all sensors. The bound $W_{i}$ is arbitrarily chosen to illustrate the results. The resulting LMI is solved using YALMIP \cite{YALMIP}, with SDP solver SDPT3 \cite{SDPT3}, providing the optimal realization $\bar{\mathcal{C}}$:
\begin{align} \vspace{-2mm} \label{eq:R_Optimal}
   \! \! \bar{\mathcal{C}} \! &\coloneqq \! \left\{ 
   \begin{aligned}
        \bar{u}_i &= \! \bar{\rho}_i \! + \!  0.771 y_{i,1} \!  - \!  0.33 y_{i,2} \\ 
        &  -0.135 y_{i,3} + 1.672 y_{i,4} + 0.187 y_{i,5}, \\
        \dot{\bar{\rho}}_i &= -0.65\bar{\rho}_i - 1.142 y_{i,1} + 0.46 y_{i,2} \\
         & \! + \! 0.222 y_{i,3} \! - \! 2.715 y_{i,4} \! - \! 0.176 y_{i,5} \! + \! 0.13 y_{i,6}.
    \end{aligned} \right.
\vspace{-2mm} \end{align}
Notably, the optimal realization $\bar{\mathcal{C}}$ maximizes robustness against $\delta_i$ by utilizing all available sensors in \eqref{eq:CR_SensorError}. 

In Fig.~\ref{fig:ReachableSet}, the outer ellipsoidal approximation of the state of \eqref{eq:CL_ReachableSet2} (using Lemma~\ref{lemma1}) is projected on the $e_i$-$z_i$ plane (using Lemma~\ref{lemma2}). Among the three controllers, despite being derived from minimizing the upper bound on the volume, $\bar{\mathcal{C}}$ has the smallest ellipsoidal approximation $(\bar{\mathcal{E}}^{\tilde{x}_i})$ in terms of volume, therefore being the most robust realization. However, the projections in Fig.~\ref{fig:ReachableSet} indicate there exists some attack where $\mathcal{C}$ is more robust than $\bar{\mathcal{C}}$, as $\bar{\mathcal{E}}^{\tilde{x}_i} \nsubseteq \mathcal{E}^{\tilde{x}_i}$.
\begin{figure}[bt]\centering
	\includegraphics[width=0.94\linewidth]{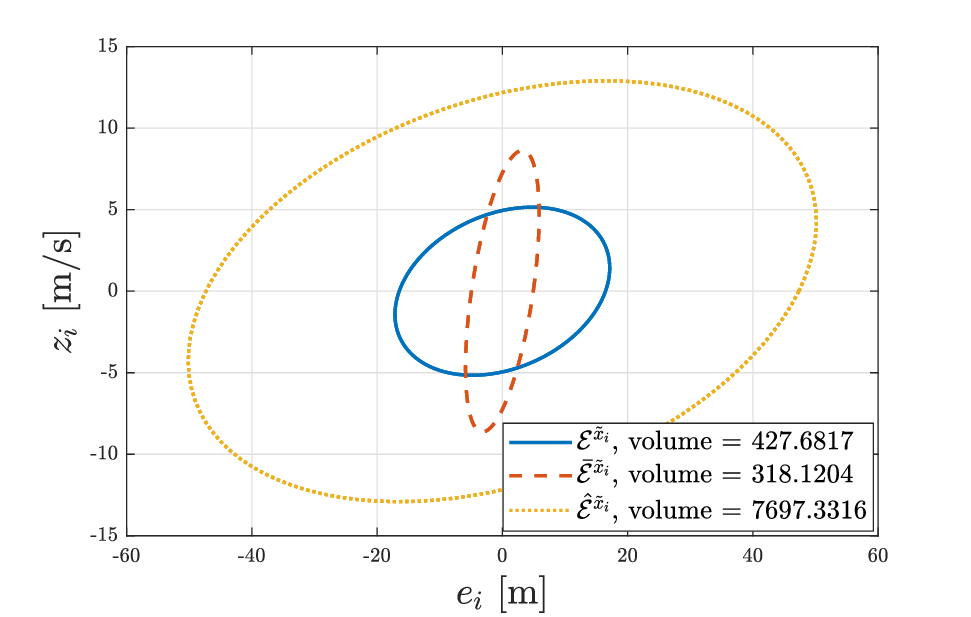}
	\vspace{-3mm}	
        \caption{Projection of the outer ellipsoidal approximation of the reachable set in \eqref{eq:CL_ReachableSet2} for the different controller realizations $\mathcal{C}$, $\bar{\mathcal{C}}$, and $\hat{\mathcal{C}}$. Results are projected onto the $z_i$-$e_i$ plane, given that $\delta_i \neq 0$, and $W_i = I$.}
	\centering
	\label{fig:ReachableSet}
        \vspace{-4mm}
\end{figure}

As a case study, all three controllers are implemented in a MATLAB Simulink simulation. The vehicles are modeled via the platooning dynamics in \eqref{eq:SD_PlatoonDynamics}, where the controllers in  \eqref{eq:SD_BaseController}, \eqref{eq:C2_Lefeber}, and \eqref{eq:R_Optimal} are used to control three different platoons. Each platoon consists of a follower (vehicle $i$), controlled via one of the realized controllers, and a leader vehicle (vehicle $i-1$), controlled via a cruise controller. The leader aims to drive a steady-state velocity of 50 km/h while accelerating and braking in the initial stage, resulting in some transient platooning behavior. In the simulation vehicle $i$ is subject to a FDI attack on $y_{i,3}$, representing the onboard acceleration measurement, where $\delta_{i,3} = \sin(3 t)\ \forall t\in [20, 75]$s. 

The results in Fig.~\ref{fig:Att_y3} show that in the initial stage the different controller realizations exhibit equivalent platooning behavior, as the platooning behavior is invariant under the coordinate transformation. During the FDI, controller $\mathcal{C}$ shows the least magnitude amplification of $\delta_{i,3}$, which corresponds with the results obtained in Fig.~\ref{fig:ReachableSet}, where $\bar{\mathcal{E}}^{\tilde{x}_i} \nsubseteq \mathcal{E}^{\tilde{x}_i}$, indicating $\bar{\mathcal{C}}$ is less robust that $\mathcal{C}$ for some FDI.
\begin{figure}[bt]\centering
    \includegraphics[width=\linewidth]{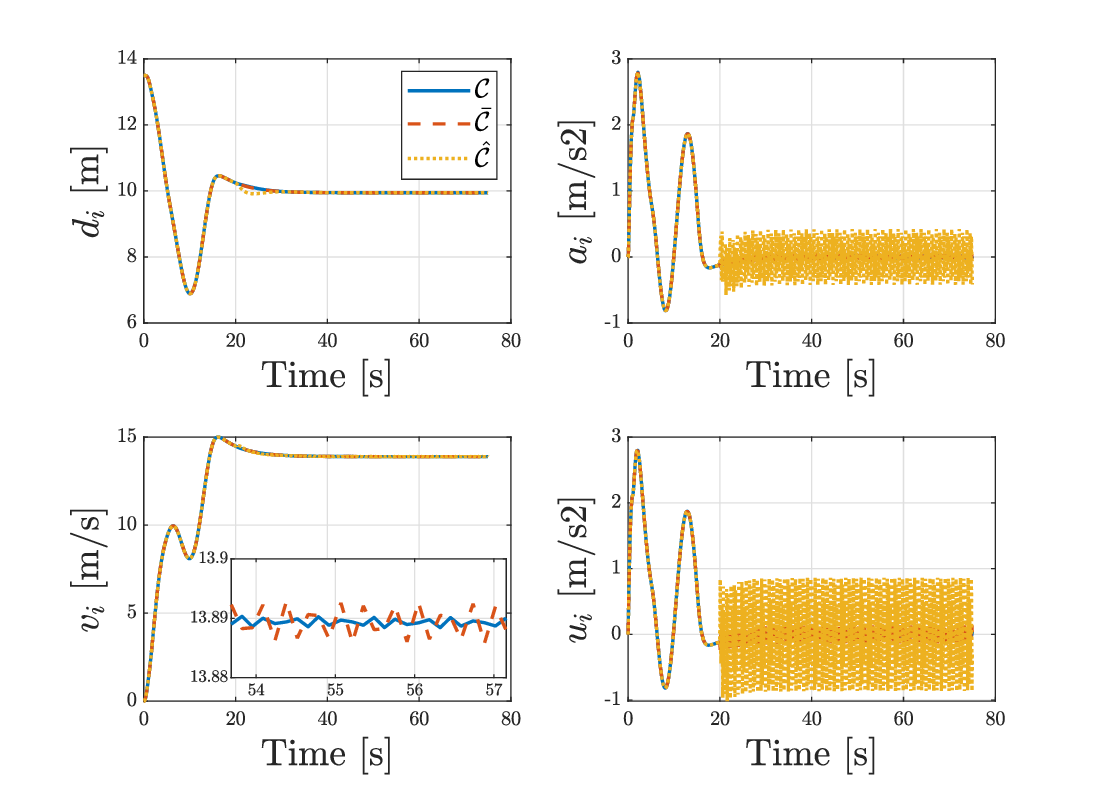}
    \vspace{-8mm}
    \caption{Vehicle and controller response for the different controller realizations $\mathcal{C}$, $\bar{\mathcal{C}}$, and $\hat{\mathcal{C}}$, given an FDI attack on $y_{i,3}$, where $\delta_{i,3} = \sin(3 t) \ \forall t \in [20, 75]$ s.}
    \centering
    \label{fig:Att_y3}
    \vspace{-6mm}
\end{figure}

}

\section{CONCLUSIONS AND FUTURE WORKS}\label{sec:Conclusion}
Cooperative driving must ensure safety and reliability in adversarial environments. To this end, we introduce the choice for a dynamic controller realization as a third controller-oriented approach to enhance the robustness of cooperative driving to cyberattacks. By reformulating a given dynamic CACC scheme, a class of equivalent controller realizations exists, having equivalent nominal behavior with varying robustness in the presence of attacks. 

Furthermore, a framework is introduced to find the optimal subset of sensors to realize the controller by minimizing the effect of FDI attacks, quantified using reachability analysis. To show our findings, three different CACC realizations are compared, one of which is the optimal realization. The optimal realization is shown to have the smallest reachable set; however, real-time simulations show that attacks still exist where different realizations are more robust. Hence, a direction of further research would be to investigate different optimization strategies to find the best realization. Another line of research concerns controller realization, considering a detection scheme that provides a more accurate bound on the FDI attacks.




\end{document}